\documentclass[fleqn,12pt,twoside]{article}
\usepackage{espcrc1}

\usepackage{graphicx}
\usepackage[figuresright]{rotating}

\newcommand{\AmS}{{\protect\the\textfont2
  A\kern-.1667em\lower.5ex\hbox{M}\kern-.125emS}}

\title{Atomic Bose-Fermi mixed condensates 
with Boson-Fermion \\ quasi-bound cluster states}

\author{H. Yabu\address[TMU]{Department of Physics, 
                             Tokyo Metropolitan University, 
                             1-1 Minami-Ohsawa, Hachioji, 
                             Tokyo 192-0397, Japan},
        Y. Takayama\addressmark[TMU],
        T. Suzuki\addressmark[TMU],
        and
        P. Schuck\address[ORS]{Institut des Sciences Nucl'eaires,Orsay,France}
}
\begin{document}

\maketitle

\begin{abstract}
The boson-fermion atomic bound states (composite fermion) 
and their roles for the phase structures 
are studied in a bose-fermi mixed condensate of atomic gas 
in finite temperature and density. 
The two-body scattering equation is formulated 
for a boson-fermion pair in the mixed condensate 
with the Yamaguchi-type potential. 
By solving the equation, 
we evaluate the binding energy of a composite fermion, 
and show that it has small $T$-dependence in the physical region, 
because of the cancellation of the boson- and fermion- statistical factors 
in the equation. 
We also calculate the phase structure
of the BF mixed condensate
under the equilibrium
$\mathrm{B}+\mathrm{F} \leftrightarrow \mathrm{BF}$, 
and discuss the role of the composite fermions:  
the competitions 
between the degenerate state of the composite fermions
and the Bose-Einstein condensate (BEC) of isolated bosons. 
The criterion for the BEC realization 
is obtained from the algebraically-derived 
phase diagrams at $T=0$.
\end{abstract}

\section{Introduction}

Since the experimental realization of the alkali-atom BEC, 
rapid progress has been made on the finite many-body physics
of quantum atomic gas--tapped atoms 
cooled down to the ultra-low temperature 
where they show quantum-mechanical behaviors.

Characteristics of the system are summarized as follows:
\begin{itemize}
\item It is a system of $10^3 \sim 10^9$ atoms 
in $T ={\rm nK} \sim {\rm \mu K}$, 
which is made by the laser-cooling technique.
The atoms are confined in a magnetically/optically-trapped potential 
of a harmonic oscillator shape (in many experiments);
a typical scale of it is given by 
$a_{{\rm HO}} \sim {\rm \mu m}$ 
in the harmonic oscillator length.
\item It is a gase of weak atomic-interaction 
in creation,  
but the interaction strength can be shifted
at any rate
by the experimental technique with Feshbach resonances 
(molecular resonances of two-body channels) 
as an external adjustable parameter. 
It is also possible to change the signature of the interaction 
(from repulsive one to attractive one and v.~v.). 
It has made possible a systematic study of many-body quantum systems 
with weak/strong-correlations.
\item A variety of precise experimental designs and observations are possible 
in the laser technology; 
for example, using quantum interference effects, 
phases of wave functions have been observed for quantum-vortex states et.al.
The time-development of systems is also traced 
(with the ${\rm n sec} \sim {\rm p sec}$ samplings), 
and has leaded to direct observations of BEC's collective oscillations.
\end{itemize}

Quantum behaviors of atomic gas in ultra-low-$T$
depend on statistical properties of the constituent atoms
(boson or fermion). 
A most interesting phenomenon in bosonic systems 
is BEC; it is the state where a large part of atoms occupies a lowest-energy 
one-body state so that their behavior is described by a wave field $\psi(x)$ 
(order parameter). 
The success of creating BEC of ${}^{87}$Rb by Colorado-JILA group in 1995
has been a break-up of the quantum atomic-gas physics, 
and a variety of phenomena has been observed on BEC \cite{rev}: 
coherence and decoherence, quantum interference, collective oscillations, 
quantum vortices, atomic lasers and so on. 

Quantum gas of fermionic atoms shows quite different behaviors.
If the atomic interaction is enough weak, 
it shows fermi-degeneracy at low-$T$; 
it was first observed by JILA group in 1999 
for the gas of ${}^{40}$K \cite{ferm}
(the first ``direct'' observation of the fermi degeneracy). 
Recently, gas of fermionic atoms is a very hot topic
in experimental and theoretical studies:
especially systems with attractive interactions, 
where possible Cooper-like-pairs or molecule formations are expected 
to make a superconducting/superfluid transition.

Another interesting system is a bose-fermi mixed condensate 
(BF mixed condensate): 
gas of two kinds of atoms, bosonic and fermionic ones, 
trapped in the same potential.
Studies by our group show 
that the interatomic interaction between bosonic and fermionic atoms 
(BF-interaction) causes many interesting phases and phenomena 
in the mixed condensates: 
collective excitations, 
induced instability \& the collapse of the system 
(Super bose nova), 
phase separations \& shell structures,
Peierls instabilities \& density waves \cite{TMU}. 
The first experiment for the BF mixed condensates 
has been done by Rice and ENS groups with ${}^{6,7}$Li isotopes \cite{bf}. 
Now, many experiments are planed to realize the BF condensates 
of a variety of atom species: ${}^{39,41-40}$K, ${}^{84,86,88-87}$Sr, 
${}^{168,170-174,176}$Yb, ${}^{87}$Rb-${}^{40}$K et.al.

Let's consider the BF mixed condensate 
with enough strongly-attractive BF-interaction.  
In such a system, 
a pair of bosonic and fermionic atoms (B and F) can be bound 
into a composite fermion (BF). 
In the physics of the mixed condensates, 
the existence of the composite fermion BF should be very important
because a group of them can give new phases of clusterized matter, e.g. 
the fermi-degenerate state of composite fermions.
In this paper, 
we discuss a structure of the BF as a bound state 
in the background of a mixed condensate in finite $T$ 
by solving a two-body scattering equation for the boson-fermion pair.
We also solve an equilibrium condition 
for the clusterization process 
${\rm B} \leftrightarrow {\rm B} +{\rm F}$, 
and obtain phase structures of the mixed condensate.
For a numerical calculation, 
we take a mixture of $^{39}$K (boson) 
and $^{40}$K (fermion) throughout this paper.

\section{Binding energy of the composite fermion in BF-mixed condensate}

\begin{figure}[htb]
\begin{minipage}[b]{90mm}
\makebox[80mm]{
\includegraphics[width=0.9\linewidth]{Fig1.EPSF}
}
\end{minipage}
\hspace{\fill}
\begin{minipage}[b]{60mm}
Figure 1. 
$T$-dependence of the binding energy of the composite fermion. BF0$\sim$BF3 corresponds to the density 
$n_b=n_f=10^{10}$, $10^{18}$, $10^{19}$, $10^{20}$atoms/cm$^3$. 
The solid line shows the $E=-T$ which provides a stability measure of the 
composite fermion at $T$.
\end{minipage}
\end{figure}

We consider a uniform gas of BF mixed condensate 
with the same mass $m$ and the BF-interaction potential $V$. 
We omit the boson-boson and fermion-fermion interactions for simplicity. 
To study a possible composite-fermion formation, 
we set up a two-body (boson-fermion) 
scattering equation in a background of the mixed condensate 
with boson- and fermion-densities $n_{B,F}$ at $T$. 
The relevant equation can be derived 
using an equation of motion \cite{RN75} 
for a boson-fermion pair operator, 
and then by replacing the average (B,F) number 
by the statistical factors $g_{B,F}(\epsilon,T,\mu_b)$  
with a kinetic energy $\epsilon$, 
where $\mu_{B,F}$ denote chemical potentials. 
The binding energy $B$ is given as a negative energy eigenvalue 
of the scattering equation. 

For the BF-interaction,
we assume a Yamaguchi type separable interaction\cite{YY54}
\begin{equation}
      \langle {\bf p}'|V|{\bf p}\rangle 
          =-\frac{\lambda}{\mu}g({\bf p}')g({\bf p}),  \qquad
     g({\bf p}) =( {\bf p}^2 +\beta^2 )^{-1},
\end{equation}
where $\bf p$ denotes a relative momentum. 
The strength $\lambda$ and the range parameter $\beta$ are in principle determined by the scattering length and the effective range of the boson-fermion scattering in free space. 
The binding energy $B$ should then be a solution of the equation
\begin{equation}
  1-4\pi\lambda\int_0^\infty d\epsilon 
  \sqrt{2m\epsilon}\left(\frac{1}{2m\epsilon+\beta^2}\right)^2
  \frac{1+g_b(\epsilon,T,\mu_b)-g_f(\epsilon,T,\mu_f)}{B+2\epsilon}=0.
\label{bfscat}
\end{equation}

As it is now possible to adjust the size and the 
sign of the atomic interaction via Feshbach resonances, 
we here concentrate on the role of the background atomic gas 
for the bound state. 
We thus fix the interaction parameters
to give a boson-fermion bound state 
in the free space, and study the solution of eq.~(\ref{bfscat}) 
as a function of the density and temperature. 

Fig.~1 shows the $T$-dependence of the BF binding energy. 
In the present formulation, 
the high-$T$ limit corresponds to the free space value. 
It is seen that the effect of the background matter 
is quite small even for a fairly low temperature; 
The deviation from the free space value is appreciable 
only below $T\sim B_{\rm free}/k_B$ and at high densities. 
This is due to a cancellation 
of the boson and fermion statistical factors in eq.~(\ref{bfscat}),
which is specific for a composite fermion. 
For the same kind of particles, 
the statistical factor comes in the equation 
with the same sign, 
and the background has much larger effect;  
For the bosonic gas, the binding energy becomes larger, 
while, for the fermionic gas, 
the Pauli blocking hinders the binding. 

\section{Phase structure of the BF-mixed condensate}

Let's consider the uniform system of polarized bosons 
\& fermions (B and F) and their composite fermions (BF), 
with the masses $m_{\mathrm{B}, \mathrm{F}, \mathrm{BF}}$.
The total number densities of B and F, 
$n_{\mathrm{Btot}, \mathrm{Ftot}}$, 
should be conserved.  

In order to obtain the phase structure at $T$ under the clusterization process
$\mathrm{B} +\mathrm{F} \leftrightarrow \mathrm{BF}$, 
we consider the equilibrium condition
\begin{equation}
     \mu_{\mathrm{B}} +\mu_{\mathrm{F}} 
     =\mu_{\mathrm{BF}} +\Delta{m} c^2,
\label{eQa}
\end{equation}
where $\mu_{\mathrm{B,F,BF}}$ are chemical potentials 
of atoms B, F, BF each other, 
and $\Delta{m} =m_{\mathrm{BF}} -m_{\mathrm{B}} -m_{\mathrm{F}}$ 
is a binding energy of the BF state. 
Because the BF mixed condensate is a weak-interacting atomic gas, 
we take the ideal-mixing approximation,  
where the chemical potentials in (\ref{eQa}) are obtained by 
the density formulae of the free bose/fermi gas:
\begin{equation}
     n_{\mathrm{B}} = \frac{(m_{\mathrm{B}})^{3/2}}{\sqrt{2} \pi^2}
          \int_0^\infty\frac{\sqrt{\epsilon}d\epsilon}{
               e^{(\epsilon-\mu_{\mathrm{B}})/k_B T}-1}, \quad
     n_a = \frac{(m_a)^{3/2}}{\sqrt{2} \pi^2}
          \int_0^\infty\frac{\sqrt{\epsilon}d\epsilon}{
               e^{(\epsilon-\mu_a)/k_B T}+1}, 
     \quad (a =\mathrm{F},\mathrm{BF}) 
      \label{eQd}
\end{equation}
where $k_B$ is a Boltzmann constant, 
and $n_{\mathrm{B,F,BF}}$ are the densities of 
the free (unpaired) B \& F and the composite BF. 

Solving eq.~(\ref{eQa}) with (\ref{eQd}) 
under the atom-number conservation for B and F: 
$n_{\mathrm{B}} +n_{\mathrm{BF}} =n_{\mathrm{Btot}}$ and
$n_{\mathrm{F}} +n_{\mathrm{BF}} =n_{\mathrm{Ftot}}$,  
we obtain the densities $n_{\mathrm{B,F,BF}}$ 
as functions of $T$ and $n_{\mathrm{Btot},\mathrm{Ftot}}$.

A special care should be paid when $T$ and $n_{\mathrm{B}}$ satisfy 
$T < T_C \equiv \frac{2\pi \hbar^2}{m_B k_B} 
\left(\frac{n_B}{2.613}\right)^{2/3}$ (BEC criterion). 
In that case, a part of free bosons condensates into BEC 
and $\mu_{\mathrm{B}}$ becomes zero;
then, the equilibrium condition 
becomes 
$\mu_{\mathrm{F}} 
=\mu_{\mathrm{BF}} +\Delta{m} c^2$.
When BEC exists, 
the condensed- and normal-component densities 
are given by 
$n_{\mathrm{BEC}} =n_{\mathrm{B}}
          \left[ 1 -\left(\frac{T}{T_C}\right)^{3/2} \right]$, 
and  
$n_{\mathrm{Bnor}} =n_{\mathrm{B}}-n_{\mathrm{BEC}}$.

\begin{figure}[htb]
\begin{minipage}[t]{80mm}
\makebox[79mm]{\includegraphics[width=0.9\linewidth]{Fig2.EPSF}}
\caption{$T$-dependence of boson density\hfil\break
in BF-mixed condensate}
\label{fig:largenenough}
\end{minipage}
\hspace{\fill}
\begin{minipage}[t]{80mm}
\makebox[79mm]{\includegraphics[width=0.9\linewidth]{Fig3.EPSF}}
\caption{Phase diagram of BF mixed condensate at $T=0\,{\rm K}$ 
         ($n_{\mathrm{Btot}} =10^{15}\,{\rm atoms/cm^3}$)}
\label{fig:toosmall}
\end{minipage}
\end{figure}

In Fig.~2,  
we show the $T$-dependence of the free boson density 
$n_{\mathrm{B}} =n_{\mathrm{BEC}}+n_{\mathrm{Bnor}}$ 
when $n_{\mathrm{Btot}} =n_{\mathrm{Ftot}} 
=10^{15}\,{\rm atoms/cm^3}$ as an typical example. 
The lines A0-A5 are for 
$\Delta{m}=(0,-3,-4,-4.71,-10) \times 10^{-6}\,{\rm K}$. 
and the oblique straight line is the critical border 
of the BEC region.
The $n_{\mathrm{B}}$ are found to decrease with decreasing $T$; 
it is because the number of composite fermions increases 
in low-$T$.
In small $\Delta{m}$ cases (A0-A3), 
the $n_{\mathrm{B}}$ is still large in low-$T$ and 
free bosons can condensate into BEC, 
but, in large $\Delta{m}$ cases (A4,A5), 
free bosons are exhausted in making composite fermions 
and the $n_{\mathrm{B}}$ becomes too small 
for the BEC realization.
The line A6 is just to the critical case. 
In high-$T$, $n_{\mathrm{B}}$ approaches to $n_{\mathrm{Btop}}$ 
in all cases; it is because all composite fermions dissociate 
into free bosons and fermions in the limit of $T \rightarrow \infty$.  

When $n_{\mathrm{Btot}} < n_{\mathrm{Ftot}}$, 
we can obtain the similar $T$-dependence in $n_{\mathrm{B}}$ as in Fig.~1,  
but, when $n_{\mathrm{Btot}} > n_{\mathrm{Ftot}}$, 
the BEC always occur in enough low-$T$
because, after all fermions are paired, 
the free bosons still remain. 

Let's consider the $T=0$ case, 
where the condition (\ref{eQa}) becomes 
$0+\epsilon_{\mathrm{F}} 
=\epsilon_{\mathrm{BF}} =\Delta{m} c^2$, 
where 
$\epsilon_a =\frac{(3\pi^2)^{2/3}}{2^{1/3} m_a} n_a^{2/3}$ 
($a=B,BF$).
We can solve this condition algebraically 
and obtain the phase structures at $T=0$. 
In Fig.~3, we show the phase diagram 
in $n_{\mathrm{Ftot}}-\Delta{m}$ plane 
in the case of $n_{\mathrm{Ftot}} =10^{15}\,{\rm atoms/cm^3}$, 
where the symbol (B,F,BF) means 
the coexistence of free bosons and 
free and composite fermions,
and so on. 
In the same figure, 
we also show phases for the $\Delta{m} > 0$ case, 
which corresponds to those with the resonance BF state.

From this diagram, 
we can read off the criterion for BEC to occur; 
it should occur in the regions when 
free bosons exist at $T=0$, e.g 
the ones with the symbol B in Fig.~3. 

\section{Summary and Discussion}

In summary, we studied bound-state structure 
of the composite fermion 
in the BF mixed condensate in finite-$T$
and showed that the $T$-dependence of the binding energy 
is shown to be small because of the cancellation 
of boson- and fermion- statistical factors in the scattering equation. 
We also obtained the phase structure of the mixed condensate 
by solving the equilibrium condition 
for the clusterization process 
$\mathrm{B}+\mathrm{F} \leftrightarrow \mathrm{BF}$. 

More details of the equilibrium calculation 
has been give in \cite{YTS}, 
and further applications 
of the present results should be discussed  
in the future publication\cite{TSYS}.

\end{document}